\documentclass[final,a4paper]{llncs}

\usepackage{etoolbox}
\newtoggle{anonymous}
\togglefalse{anonymous}

\usepackage[utf8]{inputenc}
\usepackage[utf8]{inputenc}
\usepackage{color}
\usepackage{xcolor}

\usepackage{adjustbox}

\usepackage{eucal}
\usepackage{centernot}

\usepackage{thm-restate}

\usepackage{siunitx}
\usepackage{textcomp}
\usepackage{caption}

\usepackage{pst-func}
\usepackage{amssymb}
\usepackage{graphicx}

\usepackage{paralist}
\usepackage{environ}
\usepackage{pgfplots}
\usepackage{tikz}
\usetikzlibrary{shadows,arrows,shapes,snakes,automata,backgrounds,petri}

\usepackage{centernot} % for negated arrows

\usepackage{mathtools}
\usepackage{array,multirow}

\usepackage{url}

\usepackage{pgfplotstable}
\usepackage{mdframed}
\usepackage{threeparttable}

\usepackage{etoolbox}
\usepackage{arydshln}

\usepackage{xspace} % Smart spaces after commands

\usepackage{dirtytalk} % for \say command
\usepackage{stmaryrd} % for llbracket and rrbracket\textbf{\textbf{}}

\usepackage[inline,shortlabels]{enumitem}
\newlist{inlinelist}{enumerate*}{1}
\setlist*[inlinelist,1]{%
  label=(\roman*),
}

\usepackage{xifthen}  % Extended conditional commands
\usepackage{ifthen}

\usepackage{nicefrac}

\usepackage{hyperref}
\usepackage{cleveref}
\usepackage{bigints}

\usepackage{caption,subcaption}     % Support for subfigures

\usetikzlibrary{pgfplots.dateplot}
\usetikzlibrary{matrix,chains,positioning,decorations.pathreplacing}
\usetikzlibrary{patterns,calc,fit,arrows}
\usetikzlibrary{shapes.geometric}

\usepackage[final,nomargin,inline,index]{fixme} % Simplified management of FIXME's
\fxusetheme{color}

\FXRegisterAuthor{bart}{anbart}{\color{magenta} {\underline{bart}}}
\FXRegisterAuthor{ste}{anste}{\color{blue} {\underline{ste}}}
\FXRegisterAuthor{lillo}{anlillo}{\color{orange} {\underline{lillo}}}
\FXRegisterAuthor{MM}{anMM}{\color{green} {\underline{Maurizio}}}

\hypersetup{
  breaklinks   = true,
  colorlinks   = true, %Colours links instead of ugly boxes
  urlcolor     = blue, %Colour for external hyperlinks
  linkcolor    = blue, %Colour of internal links
  citecolor    = red   %Colour of citations
}
\hypersetup{final} % Needed to generate hyperlinks even in draft mode (ERROR IN MAC)

\usepackage{listings}
\definecolor{listingBG}{HTML}{FFFFCB}%
\definecolor{listingFrame}{HTML}{BBBB98}%
\definecolor{listingLineno}{rgb}{0.5,0.5,1.0}%
\definecolor{LightGrey}{rgb}{0.975,0.975,0.975}
\lstdefinelanguage{solidity}{
	commentstyle=\color{Gray},
	morecomment=[l]{//},
	morecomment=[s]{/*}{*/},
	classoffset=0,
        escapechar=\$,
	morekeywords={struct,mapping,function,this,public,private,static,final,class,extends,switch,case,break,finally,try,catch,return,if,else,new},
	keywordstyle=\color{Blue}\bfseries,
	classoffset=1,
	morekeywords={unit,int,string,bool,address,uint},
	keywordstyle=\color{TealBlue},
	classoffset=2,
	morekeywords={ether,wei,finney,contract,send,throw,msg,sender,value},
	keywordstyle=\color{Plum}\bfseries,
}

\lstdefinelanguage{tins}{
	commentstyle=\color{Gray},
	morecomment=[l]{//},
	morecomment=[s]{/*}{*/},
	classoffset=0,
	morekeywords={skip,throw,if,then,else,while,do},
	keywordstyle=\color{Black}\bfseries,
	classoffset=1,
	morekeywords={sender,value,balance,unbound},
	keywordstyle=\valColor{}\bfseries\itshape,
	classoffset=2,
	morekeywords={contract},
	keywordstyle=\funColor{}\bfseries,
        extendedchars=true,
        literate={\$}{{\textbf{{\dollar}}}}1
}

\usepackage{tikz-cd}
\usetikzlibrary{cd,decorations.markings}
\newcommand{\ifempty}[3]{%
  \ifthenelse{\isempty{#1}}{#2}{#3}%
}

\newcommand{\ifzero}[3]{%
  \ifthenelse{\equal{#1}{0}}{#2}{#3}%
}

\newcommand{\ifdots}[3]{%
  \ifthenelse{\equal{#1}{...}}{#2}{#3}%
}

\newcommand{\hidden}[1]{}

\renewcommand{\vec}[1]{\boldsymbol{#1}}

\newcommand{\Real}[1]{\mathrm{Real}}

\newcommand{\codefont}{\fontsize{9}{9}\selectfont}
\newcommand{\code}[1]{{\tt\codefont {#1}}}

\newcommand{\codeVal}[1]{\ensuremath{\code{\valColor{#1}}}}
\newcommand{\codeFun}[1]{\ensuremath{\code{\funColor{#1}}}}

\newcommand{\bang}{{\textup{\texttt{\symbol{`\!}}}}}
\newcommand{\qmark}{{\textup{\texttt{\symbol{`\?}}}}}
\newcommand{\dollar}{{\textup{\texttt{\symbol{`\$}}}}}
\newcommand{\codeand}{\textup{\texttt{\symbol{`\&}\symbol{`\&}}}}

\def\etc{etc.\@\xspace}

\newcommand{\eg}{e.g.\@\xspace}
\newcommand{\ie}{i.e.\@\xspace}
\newcommand{\wrt}{w.r.t.\@\xspace}

  {}

\newcommand{\BTC}{\textup{%
  \leavevmode
  \vtop{\offinterlineskip %\bfseries
    \setbox0=\hbox{B}%
    \setbox2=\hbox to\wd0{\hfil\hskip-.03em
    \vrule height .3ex width .15ex\hskip .08em
    \vrule height .3ex width .15ex\hfil}
    \vbox{\copy2\box0}\box2}}\xspace}

\newcommand{\true}{\code{true}\xspace}
\newcommand{\false}{\code{false}\xspace}

\newcommand{\trueVal}{\mathit{true}\xspace}
\newcommand{\falseVal}{\mathit{false}\xspace}

\def\aColor{\color{ForestGreen}}
\def\pColor{\color{ForestGreen}}
\def\cColor{\color{ForestGreen}}

\newcommand{\Addr}{{\aColor{\textup{\textbf{Addr}}}}}
\newcommand{\Val}{{\valColor{\textup{\textbf{Val}}}}}
\newcommand{\Const}{{\valColor{\textup{\textbf{Const}}}}}

\newcommand{\aFmt}[1]{{\aColor{\mathcal{#1}}}}
\newcommand{\pFmt}[1]{{\pColor{\mathcal{#1}}}}
\newcommand{\cFmt}[1]{{\cColor{\mathcal{#1}}}}

\newcommand{\amv}[2][]{\aFmt{#2}_{\aColor{#1}}\xspace}
\newcommand{\amvA}[1][]{\amv[{#1}]{X}}
\newcommand{\amvB}[1][]{\amv[{#1}]{Y}}

\newcommand{\pmv}[2][]{\pFmt{#2}_{\pColor{#1}}\xspace}
\newcommand{\pmvA}[1][]{\pmv[{#1}]{A}}
\newcommand{\pmvB}[1][]{\pmv[{#1}]{B}}

\newcommand{\cmv}[2][]{\cFmt{#2}_{\cColor{#1}}\xspace}
\newcommand{\cmvA}[1][]{\cmv[{#1}]{C}}
\newcommand{\cmvB}[1][]{\cmv[{#1}]{D}}

\newcommand{\QmvU}[1][]{\mathbb{P}}

\definecolor{BlueViolet}{rgb}{0, 0, 0.55}
\definecolor{RubineRed}{rgb}{0.88, 0.07, 0.37}
\definecolor{ForestGreen}{rgb}{0.13, 0.55, 0.13}
\definecolor{NavyBlue}{rgb}{0.0, 0.0, 0.5}
\definecolor{Black}{rgb}{0.02, 0.02, 0.02}
\definecolor{MidnightBlue}{rgb}{0.0, 0.2, 0.4}
\definecolor{Gray}{rgb}{0.41, 0.41, 0.41}

\def\txColor{\color{MidnightBlue}}
\newcommand{\txFmt}[1]{{\txColor{\sf #1}}}

\newcommand{\tx}[2][]{\txFmt{#2}_{\txColor{#1}}}
\newcommand{\txT}[1][]{\tx[#1]{T}}

\DeclareMathAlphabet{\mathbfsf}{\encodingdefault}{\sfdefault}{bx}{n}

\newcommand{\bcEmpty}{{\mathbfsf{\txColor{\epsilon}}}}
\newcommand{\bcB}[1][]{{\boldsymbol{\mathsf{\txColor{B}}}}_{\txColor{#1}}}

\newcommand{\bcSt}[1][]{\sigma_{#1}}
\newcommand{\bcSti}[1][]{\sigma_{#1}'}
\newcommand{\bcStii}[1][]{\sigma_{#1}''}

\newcommand{\bcEnv}[1][]{\rho_{#1}}
\newcommand{\bcEnvi}[1][]{\rho_{#1}'}
\newcommand{\bcEnvii}[1][]{\rho_{#1}''}

\newcommand{\bcKV}[1][]{\delta_{#1}}

\newcommand{\mSubst}[1][]{\pi_{#1}}
\newcommand{\mSubsti}[1][]{\pi'_{#1}}
\newcommand{\irule}[2]{\dfrac{#1}{#2}}

\newcommand{\mapstopart}{\rightharpoonup}

\newcommand{\bnfdef}{::=}
\newcommand{\bnfmid}{\;|\;}

\newcommand{\nrule}[1]{{\scriptsize \textsc{#1}}}
\newcommand{\smallnrule}[1]{{\tiny \textsc{#1}}}

\newcommand{\sem}[2][]{\mbox{\ensuremath{\llbracket{#2}\rrbracket_{#1}}}}

\newcommand{\semExp}[3]{\mbox{\ensuremath{\llbracket{#1}\rrbracket_{#2}^{#3}}}}
\newcommand{\semCmd}[3]{\mbox{\ensuremath{\llbracket{#1}\rrbracket_{#2}^{#3}}}}
\newcommand{\semTx}[2]{\sem[#2]{{#1}}}
\newcommand{\semBc}[2]{\sem[#2]{{#1}}}

\newcommand{\dom}[1]{\operatorname{dom} {#1}}
\newcommand{\keys}[1]{\operatorname{keys}({#1})}

\newcommand{\Nat}{\mathbb{N}}

\newcommand{\bind}[2]{\nicefrac{#2}{#1}}
\newcommand{\setenum}[1]{\{#1\}}

\crefname{appendix}{appendix}{appendices}
\Crefname{appendix}{Appendix}{Appendices}
\crefname{notation}{notation}{notations}
\Crefname{notation}{Notation}{Notations}

\definecolor{LightGrey}{rgb}{0.95,0.95,0.95}
\definecolor{keyword}{HTML}{7F0055}

\newlength\replength
\newcommand\repfrac{.1}

\newcommand\rulewidth{.6pt}
\newcommand\tdashfill[1][\repfrac]{\cleaders\hbox to \replength{%
  \smash{\rule[\arraystretch\ht\strutbox]{\repfrac\replength}{\rulewidth}}}\hfill}

\newcommand\tdotfill[1][\repfrac]{\cleaders\hbox to \replength{%
  \smash{\raisebox{\arraystretch\dimexpr\ht\strutbox-.1ex\relax}{.}}}\hfill}

\newcommand{\var}[2][]{#2_{#1}} % variables
\newcommand{\varX}[1][]{\var[#1]{x}} 
 
\newcommand{\varY}[1][]{\var[#1]{y}}

\def\funColor{\color{RubineRed}}
\newcommand{\funFmt}[1]{{\funColor{\mathtt{#1}}}}

\newcommand{\funF}[1][]{\funFmt{f}_{\funColor{#1}}} 
 
\newcommand{\funG}[1][]{\funFmt{g}_{\funColor{#1}}} 
 
\newcommand{\funEmpty}{\funF[\cmdSkip]}

\def\valColor{\color{BlueViolet}}
\newcommand{\val}[2][]{{\valColor{#2_{#1}}}} % values
\newcommand{\valV}[1][]{\val[#1]{v}}

\newcommand{\valN}[1][]{\val[#1]{n}}
\newcommand{\valNi}[1][]{\val[#1]{n'}}
\newcommand{\valK}[1][]{\val[#1]{k}}

\def\cmdColor{\color{RubineRed}}
\newcommand{\cmdFmt}[1]{{\cmdColor{\mathit{#1}}}}
\newcommand{\cmdC}[1][]{\mathord{\cmdFmt{S}_{\cmdColor{#1}}}}
\newcommand{\cmdCi}[1][]{\mathord{\cmdFmt{S'_{#1}}}}

\newcommand{\cmdSkip}{\ensuremath{\code{skip}}}
\newcommand{\cmdAss}[2]{{#1} \code{:=} {#2}}
\newcommand{\cmdIfTE}[3]{\code{if}\, {#1}\, \code{then} \, {#2} \, \code{else} \, {#3}}
\newcommand{\cmdIfT}[2]{\code{if} \, {#1} \, \code{then} \, {#2}}
\newcommand{\cmdWhile}[2]{\code{while} \, {#1} \, \code{do} \, {#2}}
\newcommand{\cmdCall}[4][]{{#2}\ifempty{#3}{}{:{#3}({#4})}\ifempty{#1}{}{\dollar {#1}}}
\newcommand{\cmdSend}[2]{\cmdCall[#1]{#2}{}{}}
\newcommand{\cmdThrow}{\code{throw}\xspace}

\def\expColor{\color{RubineRed}}
\newcommand{\expFmt}[1]{{\cmdColor{\mathit{#1}}}}
\newcommand{\expE}[1][]{\mathord{\expFmt{E}_{\expColor{#1}}}}
\newcommand{\expEi}[1][]{\mathord{\expFmt{E'}}}
\newcommand{\expEii}[1][]{\mathord{\expFmt{E''}}}

\newcommand{\expBound}[1]{\bang{#1}}
\newcommand{\expOp}{\code{op}\xspace}
\newcommand{\semOp}{\operatorname{op}}
\newcommand{\expLookup}[1]{\qmark{#1}}

\newcommand{\expCall}[2]{{#1}:{#2}}

\newcommand{\lineno}[1]{{\tt\codefont {\textcolor{ForestGreen}{#1}}}}

\definecolor{LightGrey}{rgb}{0.95,0.95,0.95}
\definecolor{keyword}{HTML}{7F0055}

\newmdenv[linewidth=0pt]{mdNoFramed}
\makeatletter

\newcommand*{\tabminted@finalstrut}[1]{%
  \ifdim\prevdepth>0pt
    \ifdim\dp#1>\prevdepth
      \vskip\dimexpr(\dp#1)-\prevdepth\relax
    \fi
  \else
    \vskip\dimexpr(\dp#1)\relax
  \fi
}
\newcommand*{\@tabmintedend}{%
  \let\@finalstrut\tabminted@finalstrut
}
\makeatother

\newcommand{\tins}{{\textsf{TinySol}}\xspace}

\newcommand{\varBalance}{\ensuremath{\codeVal{balance}}\xspace}
\newcommand{\varSender}{\ensuremath{\codeVal{sender}}\xspace}
\newcommand{\varCaller}{\varSender}
\newcommand{\varValue}{\ensuremath{\codeVal{value}}\xspace}

\newcommand{\ethtx}[5]{{#2}\xrightarrow{#1} {#3}:{#4}({#5})}

\newcommand{\addrToContr}[1]{{\Gamma}\ifempty{#1}{}{({#1})}}
\newcommand{\contrFun}[3]{{#1}({#2}) \{ {#3} \}}
\newcommand{\contrFunSig}[2]{{#1}({#2})}

\newcommand{\mytitle}{A minimal core calculus for Solidity contracts}

\makeatletter
\renewcommand\paragraph{\@startsection{paragraph}{4}{\z@}%
  {2.25ex \@plus 1ex \@minus .2ex}%
  {-0.75em}%
  {\normalfont\normalsize\bfseries}}
\makeatother

\begin{document}

\title{\mytitle}

\iftoggle{anonymous}{}{
\author{Massimo Bartoletti\inst{1} \and Letterio Galletta\inst{2} \and Maurizio Murgia\inst{1,3}}
\authorrunning{Bartoletti et al.}

\institute{
Universit\`a degli Studi di Cagliari, Cagliari, Italy
\and
IMT Lucca, Lucca, Italy
\and
Universit\`a di Trento, Trento, Italy}
}

\maketitle

\begin{abstract}
  The Ethereum platform supports the decentralized 
  execution of \emph{smart contracts}, \ie
  computer programs that transfer digital assets between users.
  The most common language used to develop these contracts is Solidity,
  a Javascript-like language which compiles into EVM bytecode,
  the language actually executed by Ethereum nodes.
  While much research has addressed the formalisation of the semantics of 
  EVM bytecode,
  relatively little attention has been devoted to that of Solidity.
  In this paper we propose a minimal calculus for Solidity contracts,
  which extends an imperative core with a single primitive to transfer currency
  and invoke contract procedures.
  We build upon this formalisation to give semantics to the Ethereum blockchain.
  We show our calculus expressive enough to reason about 
  some typical quirks of Solidity, like \eg re-entrancy.
\end{abstract}

\keywords{Ethereum; smart contracts; Solidity}

\section{Introduction}
\label{sec:intro}

A paradigmatic feature of blockchain platforms is the ability to execute ``smart'' contracts, \ie
computer programs that transfer digital assets between users, without relying on a trusted authority.
In Ethereum~\cite{ethereum} --- the most prominent smart contracts platform so far ---
contracts can be seen as concurrent objects~\cite{Sergey17wtsc}:
they have an internal mutable state, and a set of procedures to manipulate it,
which can be concurrently called by multiple users.
Additionally, each contract controls an amount of crypto-currency, 
that it can exchange with other users and contracts.
Users interact with contracts by sending transactions,
which represent procedure calls, and may possibly
involve transfers of crypto-currency from the caller to the callee.
The sequence of transactions on the blockchain determines the
state of each contract, and the balance of each user.

Ethereum supports contracts written in a 
Turing-complete language, called EVM bytecode~\cite{Hildenbrandt18csf}.
Since programming at the bytecode level is inconvenient, 
developers seldom use EVM bytecode directly, 
but instead write contracts in higher-level languages which compile into bytecode.
The most common of these languages is Solidity~\cite{solidity},
a Javascript-like language supported by the Ethereum Foundation.
There is a growing literature on the formalization of Solidity,
which roughly can be partitioned in two approaches,
according to the distance from the formal model to the actual language.
One approach is to include as large a subset of Solidity as possible,
while the other is to devise a core calculus that is as small as possible,
capturing just the features of Solidity that are relevant to some specific task.
In general, the first approach has more direct practical applications:
for instance, a formal semantics very close to that of the actual Solidity
can be the basis for developing precise analysis and verification tools.
Although diverse in nature, the motivations underlying the second approach are not less strong.
The main benefit of omitting almost all the features of the full language is that 
by doing so we simplify rigorous reasoning.
This simplification is essential for the development of new proof techniques
(\eg, axiomatic semantics),
static analysis techniques (\eg, data and control flow analysis, type systems),
as well as for the study of language expressiveness
(\eg, rigorous comparisons and encodings to/from other languages).
The co-existence of these two approaches is common in computer science:
for instance, the formalization of Java gave rise of a lot of research since the mid 90s,
producing Featherweight Java~\cite{Igarashi01toplas}
as the most notable witness of the ``minimalistic'' approach.

\paragraph{Contribution}
In this paper we pursue the minimalistic approach,
by introducing a core calculus for smart contracts.
Our calculus, called \tins (for ``Tiny Solidity''),
features an imperative core,
which we extend with a single construct to call contracts and transfer currency.
This construct, inspired by Solidity ``external'' calls,
captures the most paradigmatic aspect of smart contracts,
\ie the exchange of digital assets according to programmable rules.
Slightly diverging from canonical presentations of imperative languages,
we use key-value stores to represent the contract state,
so abstracting and generalising Solidity state variables.
We formalise the semantics of \tins in \Cref{sec:tins:syntax}, 
using a big-step operational style.
In~\Cref{sec:tins:reentrancy}
we show \tins expressive enough to reproduce \emph{reentrancy attacks}, 
one of the typical quirks of Solidity;
the succinctness of these proofs witnesses an advantage of our minimalistic approach.
In~\Cref{sec:tins:transactions} we refine our formalization, 
by giving semantics to transactions and blockchains.
In~\Cref{sec:tins:examples} 
we exemplify \tins through a variety of complex contracts,
ranging from wallets, to escrow services, lotteries, and Ponzi schemes.
Aiming at minimality, \tins makes several simplifications \wrt Solidity:
in~\Cref{sec:conclusions} we discuss the main differences between the two languages.

\paragraph{Related work}
\label{sec:related}

Besides ours, the only other Solidity-inspired minimal core calculus
we are aware of is Featherweight Solidity (FS)~\cite{Crafa19wtsc}.
Similarly to our \tins, FS focusses on the most basic features of Solidity, 
\ie transfers of cryptocurrency and external calls,
while omitting other language features, 
like \eg internal and delegate calls, function modifiers, and the gas mechanism.
The main difference between \tins and FS is stylistic:
while our design choice was to start from a basic imperative language,
and extend it with a single contract-oriented primitive (external calls),
FS follows the style of Featherweight Java,
modelling function bodies as expressions.
Compared to our calculus, 
FS also includes the dynamic creation of contracts,
and a type system which detects some run-time errors.
A further difference is that FS models blockchains as functions
from contract identifiers to states;
instead, we represent a blockchain as a sequence of transactions,
and then we reconstruct the state by giving a semantics to this sequence.
In this way we can reason \eg about re-orderings of transactions, forks, \etc

A few papers pursue the approach of 
formalising large fragments of Solidity.
The work~\cite{Zakrzewski18vstte} proposes a big-step operational semantics for 
several Solidity constructs, including \eg 
access to memory and storage, inheritance, internal and external calls, and function modifiers.
The formalization also deals with complex data types, 
like structs, arrays and mappings. 
The works~\cite{Jiao18arxiv,Zheng18lolisa} propose
\emph{tour-de-force} formalizations of larger fragments of Solidity, 
also including a gas mechanism.
Both~\cite{Zakrzewski18vstte} and~\cite{Zheng18lolisa} mechanize their semantics in 
the Coq proof assistant,
while~\cite{Jiao18arxiv} uses the K-Framework~\cite{rosu-serbanuta-2010-jlap}.
The work~\cite{Mavridou19fc} extends the semantics of~\cite{Jiao18arxiv}
to encompass also exceptions and return values.

\section{\tins syntax and semantics}
\label{sec:tins:syntax}

We assume a set $\Val$ of \emph{values} 
$\valV, \valK, \ldots$,
a set $\Const$ of \emph{constant names} $\varX, \varY, \ldots$,
a set of \emph{procedure names} $\funF, \funG, \ldots$.
and a set $\Addr$ of \emph{addresses} $\amvA,\amvB,\ldots$,
partitioned into \emph{account addresses} $\pmvA, \pmvB, \ldots$
and \emph{contract addresses} $\cmvA, \cmvB, \ldots$.
We write sequences in bold, \eg $\vec{\valV}$ is a sequence of values; 
$\epsilon$ is the empty sequence.
We use $\valN, \valNi, \ldots$ to range over $\Nat$, 
and $b, b', \ldots$ to range over boolean values.

A \emph{contract} is a finite set of terms of the form
$\contrFun{\funF}{\vec{\varX}}{\cmdC}$, where $\cmdC$ is a \emph{statement},
with syntax in~\Cref{fig:tins:syntax}.
Intuitively, each term $\contrFun{\funF}{\vec{\varX}}{\cmdC}$ 
represents a contract procedure,
where $\funF$ is the procedure name,
$\vec{\varX}$ are its formal parameters (omitted when empty),
and $\cmdC$ is the procedure body.
Each contract has a key-value store, which we model as
a partial function from keys $\valK \in \Val$ to values $\valV \in \Val$.

Statements extend those of a basic imperative language
with three constructs: % inspired by Solidity:
\begin{itemize}[leftmargin=.15in]
\item $\cmdThrow$ raises an uncatchable exception, rolling-back the state;
\item $\cmdAss{\valK}{\expE}$ updates the key-value store,
binding the key $\valK$ to the value denoted by the expression $\expE$;
\item $\cmdCall[\valN]{\amvA}{\funF}{\vec{\valV}}$ calls the procedure $\funF$
(with actual parameters $\vec{\valV}$)
of the contract at address $\amvA$, transferring $\valN$ units of currency to $\amvA$.
\end{itemize}

The expressions used within statements (\Cref{fig:tins:syntax}, right) 
can be constants (\eg, integers, booleans, strings), 
addresses, and operations between expressions.
We assume that all the usual arithmetic, logic and cryptographic operators are provided
(since their definition is standard, we will not detail them).
The expression $\expBound{\valK}$ evaluates to $\trueVal$ if the key $\valK$
is bound in the contract store, otherwise it evaluates to $\falseVal$.
The expression $\expLookup{\valK}$ denotes the value bound
to the key $\valK$ in the contract store.
% (when $\valK$ is unbound, $\expLookup{\valK}$ evaluates to the ``undefined'' value, written $\bot$).
The expression $\expCall{\amvA}{\expE}$ evaluates $\expE$ in the context of the
address $\amvA$.
For instance, $\expCall{\amvA}{\expLookup{\valK}}$ denotes the value bound to $\valK$
in the store of $\amvA$.

We assume a mapping $\addrToContr{}$ from addresses to contracts,
such that $\addrToContr{\pmvA} = \setenum{\contrFun{\funEmpty}{}{\cmdSkip}}$ 
for all account addresses $\pmvA$.
This allows for a uniform treatment of account and contract addresses:
indeed, calling a procedure on an account address $\pmvA$
can only result in a pure currency transfer to $\pmvA$, 
since the procedure can only perform a $\cmdSkip$.
% (see~\Cref{sec:tins:semantics}).
%
We further postulate that:
\begin{inlinelist}
\item expressions and statements are well-typed:
\eg, guards in conditionals and in loops have type bool;
\item the procedures in $\addrToContr{\cmvA}$ have distinct names;
\item the key \varBalance cannot stay at the left of an assignment;
\item the constant names \varSender and \varValue 
cannot stay in the formal parameters of a procedure.
\end{inlinelist}

We use the following syntactic sugar.
For a call $\cmdCall[\valN]{\amvA}{\funF}{\vec{\valV}}$,
when there is no money transfer (\ie, $\valN=0$) we just write it as 
$\cmdCall{\amvA}{\funF}{\vec{\valV}}$;
when the target is an account address $\pmvA$
(so, the call is to the procedure $\funEmpty$),
we write it as $\cmdCall[\valN]{\pmvA}{}{}$.
We write $\cmdIfT{\expE}{\cmdC}$ for $\cmdIfTE{\expE}{\cmdC}{\cmdSkip}$.

\begin{figure}[t]
  \small
  \begin{minipage}{0.65\columnwidth}
    \begin{align*}
      \cmdC
      & \bnfdef
      && \text{statement}
      \\
      & \cmdSkip
      && \text{skip}
      \\
      \bnfmid
      & \cmdThrow
      && \text{exception}
      \\
      \bnfmid
      &  \cmdAss{\expE}{\expEi}
      && \text{store update}
      \\
      \bnfmid
      & \cmdC ; \cmdCi
      && \text{sequence}
      \\
      \bnfmid
      & \cmdIfTE{\expE}{\cmdC}{\cmdCi}
      && \text{conditional}
      \\
      \bnfmid
      & \cmdWhile{\expE}{\cmdC}
      && \text{loop}
      % \\
      % \bnfmid
      % & \cmdSend{\expE}{\expEi}
      % && \text{currency transfer}
      \\
      \bnfmid
      & \cmdCall[{\expE[2]}]{\expE[0]}{\funF}{\vec{\expE[1]}}
      && \text{call}
    \end{align*}
  \end{minipage}
  \begin{minipage}{0.315\columnwidth}
    \begin{align*}
      \expE
      & \bnfdef
      && \text{expression}
      \\
      & \valV
      && \text{value}
      \\
      \bnfmid
      & \varX
      && \text{const name}
      \\
      \bnfmid
      & \amvA
      && \text{address}
      \\
      \bnfmid
      & \expOp \; \vec{\expE}\;
      && \text{operator}
      \\
      \bnfmid
      & \expLookup{\expE}
      && \text{key lookup}
      \\
      \bnfmid 
      & \expBound{\expE}
      && \text{key bound?}
      \\
      \bnfmid
      & \expCall{\amvA}{\expE}
      && \text{context}
    \end{align*}
  \end{minipage}
  \caption{Syntax of \tins.}
  \label{fig:tins:syntax}
\end{figure}

The semantics of contracts is given in terms of a function from 
states to states.
A \emph{state} $\bcSt : \Addr \rightarrow (\Val \mapstopart \Val)$
maps each address to a key-value store,
\ie a partial function from values (keys) to values.
We use the standard brackets notation for representing finite maps: 
for instance, 
$\setenum{\bind{\varX[1]}{\valV[1]},\cdots,\bind{\varX[n]}{\valV[n]}}$
maps $\varX[i]$ to~$\valV[i]$, for $i \in 1..n$.
When a key $\valK$ is not bound to any value in $\bcSt \, \amvA$,
we write $\bcSt \, \amvA \, \valK = \bot$.
We postulate that $\dom{\bcSt \pmvA} = \setenum{\varBalance}$
for all account addresses $\pmvA$, and
$\dom{\bcSt \cmvA} \supseteq \setenum{\varBalance}$
for all contract addresses $\cmvA$.
A \emph{qualified key} is a term of the form $\amvA.\valK$.
We write $\bcSt (\amvA.\valK)$ for $\bcSt \amvA \valK$.

A \emph{state update} 
$\mSubst : \Addr \mapstopart (\Val \mapstopart \Val)$
is a substitution from qualified keys to values;
we denote with $\setenum{\bind{\amvA.\valK}{\valV}}$
the state update which maps $\amvA.\valK$ to~$\valV$.
We define $\keys{\mSubst}$ as the set of qualified keys $\amvA.\valK$ such that
$\amvA \in \dom{\mSubst}$ and $\valK \in \dom{\mSubst \amvA}$.
We apply updates to states as follows:
\[
(\bcSt \mSubst) \amvA = \bcKV[\amvA]
\quad \text{where} \quad
\bcKV[\amvA]  \valK =
\begin{cases}
  \mSubst \amvA \valK & \text{if $\amvA.\valK \in \keys{\mSubst}$} \\
  \bcSt \amvA \valK & \text{otherwise}
\end{cases}
\]
We define the auxiliary operators $\bcSt + \amvA:\valN$ and 
$\bcSt - \amvA:\valN$ on states,
which, respectively, increase/decrease
the $\varBalance$ of $\amvA$ of $\valN$ currency units:
\[
\bcSt \circ \amvA:\valN
\; = \;
\bcSt\setenum{\bind{\amvA.\varBalance}{(\bcSt \amvA \varBalance) \,\circ\, \valN}}
\tag*{($\circ \in \setenum{+,-}$)}
\]

\begin{example}
  Let $\bcSt$ be a state which maps $\amvA$ to the store 
  $\bcKV[\amvA] = \setenum{\bind{\valK[0]}{0},\bind{\valK[1]}{1}}$.
  Let $\mSubst = \setenum{\bind{\amvA.\valK[0]}{2}}$ and
  $\mSubsti = \setenum{\bind{\amvA.\valK[2]}{3}}$
  be state updates.
  We have
  $(\bcSt \mSubst) \amvA = \setenum{\bind{\valK[0]}{2},\bind{\valK[1]}{1}}$,
  $(\bcSt \mSubsti) \amvA = \setenum{\bind{\valK[0]}{0},\bind{\valK[1]}{1},\bind{\valK[2]}{3}}$, 
  and $(\bcSt \mSubst) \amvB  = (\bcSt \mSubsti) \amvB = \bcSt \amvB$
  for all $\amvB \neq \amvA$.
\end{example}

We give the operational semantics of statements in a big-step style.
The semantics of a statement $\cmdC$ is parameterised over
a state $\bcSt$, an address $\amvA$ 
(the contract wherein $\cmdC$ is evaluated),
and an \emph{environment} $\bcEnv : \Const \mapstopart \Val$,
used to evaluate the formal parameters
and the special names $\varSender$ and $\varValue$.

\begin{figure*}[t]
  \resizebox{\textwidth}{!}{
    \begin{minipage}{1.05\textwidth}
      \[
      \begin{array}[c]{c}
        \semExp{\valV}{\bcSt,\bcEnv}{\amvA} = \valV
        \quad\;\;
        \semExp{\varX}{\bcSt,\bcEnv}{\amvA} = \bcEnv \, \varX
        \quad\;\;
        \semExp{\amvB}{\bcSt,\bcEnv}{\amvA} = \amvB
        \quad\;\;
        {\semExp{\expOp \vec{\expE}}{\bcSt,\bcEnv}{\amvA} = \semOp \semExp{\vec{\expE}}{\bcSt,\bcEnv}{\amvA}}
        \quad\;\;
        {\semExp{\expCall{\amvB}{\expE}}{\bcSt,\bcEnv}{\amvA}
        =
        \semExp{\expE}{\bcSt,\bcEnv}{\amvB}}
        \\[10pt]
        {\semExp{\expLookup{\expE}}{\bcSt,\bcEnv}{\amvA}
        =
        \bcSt \, \amvA \, (\semExp{\expE}{\bcSt,\bcEnv}{\amvA})}
        \qquad
        \semExp{\expBound \expE}{\bcSt,\bcEnv}{\amvA} = 
        \begin{cases}
          \trueVal  & \text{if $\semExp{\expE}{\bcSt,\bcEnv}{\amvA} \neq \bot$ and $\bcSt \, \amvA \, ( \semExp{\expE}{\bcSt,\bcEnv}{\amvA}) \neq \bot$} \\
          \falseVal & \text{if $\semExp{\expE}{\bcSt,\bcEnv}{\amvA} \neq \bot$ and $\bcSt \, \amvA \, ( \semExp{\expE}{\bcSt,\bcEnv}{\amvA}) = \bot$}
        \end{cases}
        \\[15pt]
        %%% skip
        \irule
        {}
        {\semCmd{\cmdSkip}{\bcSt,\bcEnv}{\amvA} = \bcSt}
        \quad
        %%% store update
        \irule
        {\semExp{\expE}{\bcSt,\bcEnv}{\amvA} = \valK \quad
        \semExp{\expEi}{\bcSt,\bcEnv}{\amvA} = \valV}
        {\semCmd{\cmdAss{\expE}{\expEi}}{\bcSt,\bcEnv}{\amvA} =
        \bcSt\setenum{\bind{\amvA.\valK}{\valV}}}
        \quad
        %%% conditional
        \irule
        {\semExp{\expE}{\bcSt,\bcEnv}{\amvA} = b \in \setenum{\trueVal,\falseVal}}
        {\semCmd{\cmdIfTE{\expE}{\cmdC[\trueVal]}{\cmdC[\falseVal]}}{\bcSt,\bcEnv}{\amvA} =
        \semCmd{\cmdC[b]}{\bcSt,\bcEnv}{\amvA}}
        \\[15pt]
        %%% sequence
        \irule
        {\semCmd{\cmdC[0]}{\bcSt,\bcEnv}{\amvA} = \bcSti}
        {\semCmd{\cmdC[0]; \cmdC[1]}{\bcSt,\bcEnv}{\amvA} = \semCmd{\cmdC[1]}{\bcSti,\bcEnv}{\amvA}}
        \quad
        %%% while (false)
        \irule
        {\semExp{\expE}{\bcSt,\bcEnv}{\amvA} = \falseVal}
        {\semCmd{\cmdWhile{\expE}{\cmdC}}{\bcSt,\bcEnv}{\amvA} = \bcSt}
        \quad
        %%% while (true)
        \irule
        {\semExp{\expE}{\bcSt,\bcEnv}{\amvA} = \trueVal \quad
        \semCmd{\cmdC}{\bcSt,\bcEnv}{\amvA} = \bcSti}
        {\semCmd{\cmdWhile{\expE}{\cmdC}}{\bcSt,\bcEnv}{\amvA} =
        \semCmd{\cmdWhile{\expE}{\cmdC}}{\bcSti,\bcEnv}{\amvA}}
        \\[15pt]
        %%% procedure call
        \irule
        {\begin{array}{l}
           \semExp{\expE[0]}{\bcSt,\bcEnv}{\amvA} = \amvB
           \\
           \semExp{\vec{\expE[1]}}{\bcSt,\bcEnv}{\amvA} = \valV[1] \cdots \valV[h]
         \end{array}
        \quad
        \begin{array}{l}
          \semExp{\expE[2]}{\bcSt,\bcEnv}{\amvA} = \valN \leq \bcSt \, \amvA \, \varBalance
          \\
          \contrFun{\funF}{\varX[1] \cdots \varX[h]}{\cmdC} \in \addrToContr{\amvB}
        \end{array}
        \quad
        \begin{array}{l}
          \bcSti = \bcSt - \amvA:\valN + \amvB:\valN
          \\
          \bcEnvi = \setenum{\bind{\varSender}{\amvA},\bind{\varValue}{\valN},\bind{\varX[1]}{\valV[1]}, \cdots, \bind{\varX[h]}{\valV[h]}}
        \end{array}
        }
        {\semCmd{\cmdCall[{\expE[2]}]{\expE[0]}{\funF}{\vec{\expE[1]}}}{\bcSt,\bcEnv}{\amvA} =
        \semCmd{\cmdC}{\bcSti, \, \bcEnvi}{\amvB}}
      \end{array}
      \]
    \end{minipage}
  }
  \caption{Semantics of statements and expressions.}
  \label{fig:tins:semantics}
\end{figure*}

Executing $\cmdC$ may affect both the store of $\amvA$ and,
in case of procedure calls, also the store of other contracts.
Instead, the semantics of an expression 
is a value; so, expressions have no side effects.
We assume that all the semantic operators are \emph{strict},
\ie their result is $\bot$ if some operand is $\bot$.
% (\eg, $\bcSt \amvA \bot = \bot$).
%
We denote by $\semCmd{\cmdC}{\bcSt,\bcEnv}{\amvA}$ 
the semantics of a statement $\cmdC$  
in a given state $\bcSt$, environment $\bcEnv$, and address $\amvA$,
where the partial function $\semCmd{\cdot}{\bcSt,\bcEnv}{\amvA}$ is defined
by the inference rules in~\Cref{fig:tins:semantics}.
We write $\semCmd{\cmdC}{\bcSt,\bcEnv}{\amvA} = \bot$ 
when the semantics of $\cmdC$ is not defined.

The semantics of expressions is straightforward;
note that we use $\expOp$ to denote syntactic operators,
and $\semOp$ for their semantic counterpart.
The environment $\bcEnv$ is used to evaluate constant names $\varX$,
while the state $\bcSt$ is used to evaluate 
$\expBound{\expE}$ and $\expLookup{\expE}$.
The semantics of statements is mostly standard, except for the last rule.
A procedure call $\cmdCall[{\expE[2]}]{\expE[0]}{\funF}{\vec{\expE[1]}}$ 
within $\amvA$ 
has a defined semantics iff:
\begin{inlinelist}
\item $\expE[0]$ evaluates to an address $\amvB$;
\item $\expE[2]$ evaluates to a non-negative number $\valN$,
not exceeding the balance of $\amvA$;
\item the contract at $\amvB$ has a procedure named $\funF$ 
with formal parameters $\varX[1] \cdots \varX[h]$;
\item $\vec{\expE[1]}$ evaluates to a sequence of values of length $h$.
\end{inlinelist}
If all these conditions hold, then
the procedure body $\cmdC$ is executed
in a state where 
$\amvA$'s balance is decreased by $\valN$, 
$\amvB$'s balance is increased by $\valN$,
and in an environment where the formal parameters are bound to the actual ones,
and the special names \varSender and \varValue are bound, respectively, 
to $\amvA$ (the caller) and $\valN$ (the value transferred to $\amvB$).

% https://solidity.readthedocs.io/en/v0.5.3/units-and-global-variables.html

\begin{example}
  \label{ex:tins:bot}
  Consider the following statements, to be evaluated within a contract $\cmvA$
  in a store $\bcSt$ where $\bcSt \cmvA \valK = \bot$:
  \begin{align*}
    & \cmdAss{\expLookup{\valK}}{1}
    && \cmdAss{\valK}{\expLookup{\valK}}
    && \cmdIfTE{\expBound{\expLookup{\valK}}}{\cmdAss{\valK}{0}}{\cmdAss{\valK}{1}}
    \\
    & \cmdThrow
    && \cmdAss{\expLookup{\valK}}{1} ; \; \cmdSkip
    % && \cmdAss{\valK}{0} ; \; \cmdThrow
    && \cmdWhile{\true}{\cmdSkip}
  \end{align*}
  % $\semCmd{\cmdC}{\bcSt,\bcEnv}{\amvA} = \bot$,
  We have that:
  \begin{inlinelist}[(a)]
    
  \item $\cmdAss{\expLookup{\valK}}{1}$ evaluates to $\bot$
    because the first premise of the assignment rule is not satisfied,
    as the lhs of the assignment evaluates to $\bot$;

  \item similarly, $\cmdAss{\valK}{\expLookup{\valK}}$ evaluates to $\bot$
    because the second premise is not satisfied,
    as the rhs evaluates to $\bot$;
  
  \item $\cmdIfTE{\expBound{\expLookup{\valK}}}{\cmdAss{\valK}{0}}{\cmdAss{\valK}{1}}$ evaluates to $\bot$,
    because the semantics of the guard is $\bot$;

  \item since there are no semantic rules for $\cmdThrow$,
    implicitly this means that its semantics is undefined;

  \item $\cmdAss{\expLookup{\valK}}{1} ; \; \cmdSkip$
    is a sequence of two commands, where the first command evaluates to $\bot$.
    The rule for sequences requires that the first command evaluates to some state $\bcSti$, while this is not the case for 
    $\cmdAss{\expLookup{\valK}}{1}$.
    Therefore, the premise of the rule does not hold, 
    and so the overall command evaluates to $\bot$;

  \item finally, $\cmdWhile{\true}{\cmdSkip}$ evaluates to $\bot$, 
    because there exists no
    (finite) derivation tree which infers 
    $\semCmd{\cmdWhile{\true}{\cmdSkip}}{\bcSt,\bcEnv}{\cmvA} = \bcSt$.

  \end{inlinelist}
  Summing up, all the statements above have an undefined semantics.
  In practice, 
  the semantic rules for transactions (see \Cref{sec:tins:transactions}) 
  ensure that the effects of any transaction whose statement evaluates 
  to $\bot$ will be reverted
  (see \eg \Cref{ex:tins:wallet:transactions}).
\end{example}

\begin{example}[Wallet]
  \label{ex:tins:wallet}
  % f() { if sender=A then skip else throw } 
  % g(x,y) { if sender=A and ?balance>x then y $ x else throw } 
  Consider the following procedures of the contract at $\cmvA$:
  \begin{align*}
    \contrFunSig{\funF}{} \, \{ 
    & \cmdIfTE{\varSender = \pmvA}{\cmdSkip}{\cmdThrow} \}
    \\
    \contrFunSig{\funG}{\varX,\varY} \, \{ 
    & \code{if} \; {\varSender = \pmvA  \,\codeand\, \varValue=0 \,\codeand\, \expLookup{\varBalance}\geq\varX}
      \;\code{then}\; {\cmdSend{\varX}{\varY}} \; \code{else} \; \cmdThrow \}
  \end{align*}
  The procedure $\funF$ allows $\pmvA$ to deposit funds to the contract;
  dually, $\funG$ allows $\pmvA$ to transfer funds to other addresses.
  The guard $\varSender = \pmvA$ ensures that only $\pmvA$ can invoke the procedures of $\cmvA$;
  calls from other addresses result in a $\cmdThrow$, 
  which leaves the state of $\cmvA$ unchanged
  (in particular, $\cmdThrow$ reverts the currency transfer from $\varSender$ to $\cmvA$).
  The procedure $\funG$ also checks that no currency 
  is transferred along with the contract call
  ($\varValue=0$),
  and that the balance of $\cmvA$ is enough ($\expLookup{\varBalance}\geq\varX$).
  % Let $\cmdC[\funF] = \cmdIfTE{\varSender = \pmvA}{\cmdSkip}{\cmdThrow}$, and let
  Let $\cmdC[\funG]$ be the body of $\funG$,  
  let $\bcSt$ be such that $\bcSt \cmvA \varBalance = 3$,
  and let $\bcEnv = \setenum{\bind{\varSender}{\pmvA},\bind{\varValue}{0},\bind{\varX}{2},\bind{\varY}{\pmvB}}$.
  We have:
  \begin{align*}
    \semCmd{\cmdC[\funG]}{\bcSt,\bcEnv}{\cmvA}    
    & = \semCmd{\cmdSend{\varX}{\varY}}{\bcSt,\bcEnv}{\cmvA}
      \; = \;
      \semCmd{\cmdCall[\varX]{\varY}{\funEmpty}{}}{\bcSt,\bcEnv}{\cmvA}
    = \semCmd{\cmdSkip}{\bcSt - \cmvA:2 + \pmvB:2,\, \setenum{\bind{\varCaller}{\cmvA},\bind{\varValue}{1}}}{\pmvB}
    \\
    & = \bcSt - \cmvA:2 + \pmvB:2
  \end{align*}
  Note that $\semCmd{\cmdC[\funG]}{\bcSt,\bcEnv}{\cmvA} = \bot$
  if $\bcSt \cmvA \varBalance < 2$, or
  $\bcEnv \varSender \neq \pmvA$, or $\bcEnv \varValue \neq 0$.
\end{example}

\section{Digression: modelling re-entrancy}
\label{sec:tins:reentrancy}

We now show how to express in \tins \emph{re-entrancy}, 
a subtle features of Solidity which was exploited in the famous 
``DAO Attack''~\cite{ABC17post,Luu16ccs}.

\begin{example}[Harmless re-entrancy]
  \label{ex:tins:reentrancy}
  Consider the following procedures:
  %\begin{small}
    \begin{align*}
      & \contrFun{\funF}{x,b}{\cmdIfT{b}{\{ 
        \cmdCall{\cmvB}{\funG}{}; \; 
        \cmdSend{\varValue}{x} \} }}
        % {\cmdSkip}}
      && \in \addrToContr{\cmvA}
      \\
      & \contrFun{\funG}{}{\cmdCall{\varSender}{\funF}{\pmvB,\false}}
      && \in \addrToContr{\cmvB}
    \end{align*}%
  %\end{small}%
  %
  Intuitively, $\funF$ first calls $\funG$, and then transfers $\varValue$ units of currency
  to the address $\varX$.
  The procedure $\funG$ attempts to change the currency recipient
  by calling back $\funF$, setting the parameter $\varX$ to $\pmvB$.
  We prove that this attack fails.
  Let $\cmdC = \cmdCall[1]{\cmvA}{\funF}{\pmvA,\true}$.
  For all $\bcSt$ and $\bcEnv$ such that $\bcSt \cmvA \varBalance = 1$,
  we have:
  \begin{small}
    \begin{align*}
      \semCmd{\cmdC}{\bcSt,\bcEnv}{\amvA}
      & = \semCmd{\cmdIfT{b}{\{ 
        \cmdCall{\cmvB}{\funG}{}; \; 
        \cmdSend{\varValue}{x} \} }}
        % {\cmdSkip}}
        {\bcSt,\bcEnvi}{\cmvA}
      && (\bcEnvi = \setenum{\bind{\varSender}{\cmvA},\bind{\varValue}{1},\bind{b}{\trueVal},\bind{\varX}{\pmvA}})
      \\
      & = \semCmd{ 
        \cmdCall{\cmvB}{\funG}{}; \; 
        \cmdSend{\varValue}{x}}{\bcSt,\bcEnvi}{\cmvA}
      \\
      & = \semCmd{ 
        \cmdSend{\varValue}{x}}{\bcSti,\bcEnvi}{\cmvA}
      && (\bcSti = \semCmd{\cmdCall{\cmvB}{\funG}{}}{\bcSt,\bcEnvi}{\cmvA})
      \\
      & = \bcSti - \cmvA:1 + \pmvA:1
      \\
      \text{where } \bcSti 
      & = \semCmd{\cmdCall{\cmvB}{\funG}{}}{\bcSt,\bcEnvi}{\cmvA}
      \\
      & = \semCmd{\cmdCall{\varSender}{\funF}{\pmvB,\false}}{\bcSt,\setenum{\bind{\varSender}{\cmvA},\bind{\varValue}{0}}}{\cmvB}
      \\
      & = \semCmd{
        \cmdIfT{b}{\{ 
        \cmdCall{\cmvB}{\funG}{}; \; 
        \cmdSend{\varValue}{x} \} }}
        % {\cmdSkip}}
        {\bcSt,\bcEnvii}{\cmvA}
      && (\bcEnvii = \setenum{\bind{\varSender}{\cmvB},\bind{\varValue}{0},\bind{b}{\falseVal},\bind{\varX}{\pmvB}})
      \\
      & = \semCmd{\cmdSkip}
        {\bcSt,\bcEnvii}{\cmvA}
        = \bcSt
    \end{align*}%
  \end{small}%
  Since $\bcSti = \bcSt$, we conclude that
  $\semCmd{\cmdC}{\bcSt,\bcEnv}{\cmvA} = \bcSt - \cmvA:1 + \pmvA:1$.
  So, $\funG$ has failed its attempt to divert the currency transfer to~$\pmvB$.
  \qed
\end{example}

\begin{example}[Vicious re-entrancy]
  \label{ex:tins:vicious}
  Consider the following procedures:
  % \begin{small}
    \begin{align*}
      \contrFunSig{\funF}{} \{ 
      & \code{if} \; \code{not}\expBound{\valK} \,\codeand\, \expLookup{\varBalance} \geq 1 
        \; \code{then} \; \{ \cmdCall[1]{\cmvB}{\funG}{}; \; \cmdAss{\valK}{\true} \} \; 
        % \code{else} \; {\cmdSkip} 
        \}
      && \in \addrToContr{\cmvA}
      \\
      \contrFunSig{\funG}{} \{ 
      & \cmdCall{\cmvA}{\funF}{} \}
      && \in \addrToContr{\cmvB}
    \end{align*}%
  % \end{small}%
  % 
  Intuitively, $\funF$ would like to transfer $1$ ether to $\cmvB$, 
  by calling $\funG$.
  The guard $\code{not}\expBound{\valK}$ is intended to ensure that the transfer
  happens at most once.
  Let $\bcSt$ be such that $\bcSt \cmvA \varBalance = \valN \geq 1$
  and $\bcSt \cmvA \valK = \bot$,
  and let $\bcEnv = \setenum{\bind{\varSender}{\cmvB},\bind{\varValue}{0}}$,
  $\bcEnvi = \setenum{\bind{\varSender}{\cmvA},\bind{\varValue}{1}}$.
  Let $\cmdC[\funF]$ and $\cmdC[\funG]$ be the bodies of $\funF$ and $\funG$.
  We have:
  \begin{small}
    \begin{align*}
      \semCmd{\cmdC[\funF]}{\bcSt,\bcEnv}{\cmvA}
      & = \semCmd{\cmdCall[1]{\cmvB}{\funG}{}; \; \cmdAss{\valK}{\true}}{\bcSt,\bcEnv}{\cmvA}
        = \semCmd{\cmdAss{\valK}{\true}}{\bcSt[1],\bcEnv}{\cmvA}
      \\
      \bcSt[1]
      &  = \semCmd{\cmdCall[1]{\cmvB}{\funG}{}}{\bcSt,\bcEnv}{\cmvA}
        = \semCmd{\cmdC[\funG]}{\bcSt - \cmvA:1 + \cmvB:1, \bcEnvi}{\cmvB}
        = \semCmd{\cmdC[\funF]}{\bcSt - \cmvA:1 + \cmvB:1, \bcEnv}{\cmvA}
      \\
      & = \semCmd{\cmdCall[1]{\cmvB}{\funG}{}; \; \cmdAss{\valK}{\true}}{\bcSt - \cmvA:1 + \cmvB:1, \bcEnv}{\cmvA}
      \\
      & = \semCmd{\cmdAss{\valK}{\true}}{\bcSt[2],\bcEnv}{\cmvA}
      \\
      \bcSt[2]
      &  = \semCmd{\cmdCall[1]{\cmvB}{\funG}{}}{\bcSt  - \cmvA:1 + \cmvB:1,\bcEnv}{\cmvA}
        = \semCmd{\cmdC[\funG]}{\bcSt - \cmvA:2 + \cmvB:2, \bcEnvi}{\cmvB}
        = \semCmd{\cmdAss{\valK}{\true}}{\bcSt[3],\bcEnv}{\cmvA}
      \\
      \bcSt[i] & = \semCmd{\cmdAss{\valK}{\true}}{\bcSt[i+1],\bcEnv}{\cmvA} \qquad\qquad \text{(for $i \in 3 \ldots n-1$)}
      \\
      \bcSt[n] & = \semCmd{\cmdSkip}{\bcSt - \cmvA:n + \cmvB:n,\bcEnv}{\cmvA}
                 = \bcSt - \cmvA:n + \cmvB:n
    \end{align*}%
  \end{small}%
  Summing up, $\semCmd{\cmdC[\funF]}{\bcSt,\bcEnv}{\cmvA} = (\bcSt - \cmvA:n + \cmvB:n) \setenum{\bind{\valK}{\trueVal}}$,
  \ie $\cmvB$ has drained all the currency from $\cmvA$.
  \qed
\end{example}

\section{Transactions and blockchains}
\label{sec:tins:transactions}

A \emph{transaction} $\txT$ is a term of the form
\(
\ethtx{\valN}{\pmvA}{\cmvA}{\funF}{\vec{\valV}}
\),
where $\pmvA$ is the address of the caller,
$\cmvA$ is the address of the called contract,
$\funF$ is the called procedure,
$n$ is the value transferred from $\pmvA$ to $\cmvA$, and
$\vec{\valV}$ is the sequence of actual parameters.
The semantics of $\txT$ in a given state $\bcSt$, 
is a new state $\bcSti = \semTx{\txT}{\bcSt}$.
The function $\semTx{\cdot}{\bcSt}$ is defined by the following rules:
\[
\small
\begin{array}{c}
  \irule
  {\contrFun{\funF}{\vec{\varX}}{\cmdC} \in \addrToContr{\cmvA}
  \qquad
  \bcSt \, \pmvA \, \varBalance \geq \valN
  \qquad
  \semCmd{\cmdC}{\bcSt-\pmvA:\valN+\cmvA:\valN,\, \setenum{\bind{\varSender}{\pmvA},\bind{\varValue}{\valN},\bind{\vec{\varX}}{\vec{\valV}}}}{\cmvA} = \bcSti
  }
  {\semTx{\ethtx{\valN}{\pmvA}{\cmvA}{\funF}{\vec{\valV}}}{\bcSt} = \bcSti}
  \smallnrule{[Tx1]}
  \\[15pt]
  \irule
  {\contrFun{\funF}{\vec{\varX}}{\cmdC} \in \addrToContr{\cmvA}
  \;\;
  \big(
  \bcSt \, \pmvA \, \varBalance < \valN \quad \text{or} \quad
  \!\semCmd{\cmdC}{\bcSt-\pmvA:\valN+\cmvA:\valN,\, \setenum{\bind{\varSender}{\pmvA},\bind{\varValue}{\valN},\bind{\vec{\varX}}{\vec{\valV}}} }{\cmvA} = \bot
  \big)
  }
  {\semTx{\ethtx{\valN}{\pmvA}{\cmvA}{\funF}{\vec{\valV}}}{\bcSt} = \bcSt}
  \smallnrule{[Tx2]}
\end{array}
\]

Rule~\nrule{[Tx1]} handles the case where the transaction is successful:
this happens when $\pmvA$'s balance is at least $\valN$,
and the procedure call terminates in a non-error state.
Note that $\valN$ units of currency are transferred to $\cmvA$
\emph{before} starting to execute $\funF$,
and that the names $\varSender$ and $\varValue$ are set, respectively,
to $\pmvA$ and $\valN$.
Instead, \nrule{[Tx2]} applies either when $\pmvA$'s balance is not enough,
or the execution of $\funF$ fails (this also covers the case when $\funF$ does not terminate).
In these cases, $\txT$ does not alter the state, \ie $\bcSti = \bcSt$.

A \emph{blockchain} $\bcB$ is a finite sequence of transactions.
The semantics of $\bcB$ is obtained by folding 
the semantics of its transactions:
\[
\semBc{\bcEmpty}{\bcSt} = \bcSt
\qquad
\semBc{\txT \bcB}{\bcSt} = \semBc{\bcB}{\scriptsize \semTx{\txT}{\bcSt}}
\]
Note that erroneous transactions occuring in a blockchain
have no effect on its semantics (as rule \nrule{[Tx2]} makes them identities \wrt the append operation).

\begin{example}
  \label{ex:tins:wallet:transactions}
  Recall the contract $\cmvA$ from~\Cref{ex:tins:wallet}, and let
  $\bcB = \txT[0] \txT[1] \txT[0]$, where:
  \[
  \txT[0] = \ethtx{3}{\pmvA}{\cmvA}{\funF}{}
  \qquad
  \txT[1] = \ethtx{0}{\pmvA}{\cmvA}{\funG}{2,\pmvB}
  \]
  Let $\cmdC[\funF]$ and $\cmdC[\funG]$ be the bodies of $\funF$ and $\funG$,
  respectively.
  % = \cmdIfTE{\varSender = \pmvA}{\cmdSkip}{\cmdThrow}$  Let $\bcSt$ be such that
  $\bcSt \pmvA \varBalance = 5$ and
  $\bcSt \cmvA \varBalance = 0$.
  By rule \nrule{[Tx1]} we have that:
  \begin{align*}
    \semTx{\txT[0]}{\bcSt}
    & = \semCmd{\cmdC[\funF]}{\bcSt - \pmvA:3 + \cmvA:3, \setenum{\bind{\varCaller}{\pmvA},\bind{\varValue}{3}}}{\cmvA}
      = \semCmd{\cmdSkip}{\bcSt - \pmvA:3 + \cmvA:3, \setenum{\bind{\varCaller}{\pmvA},\bind{\varValue}{3}}}{\cmvA}
    \\
    & = \bcSt - \pmvA:3 + \cmvA:3
  \end{align*}
  Now, let $\bcSti = \bcSt - \pmvA:3 + \cmvA:3$.
  By rule \nrule{[Tx1]} we have that:
  \begin{align*}
    \semTx{\txT[1]}{\bcSti}
    & = \semCmd{\cmdC[\funG]}{\bcSti, \setenum{\bind{\varCaller}{\pmvA},\bind{\varValue}{0},\bind{\varX}{2},\bind{\varY}{\pmvB}}}{\cmvA}
      = \semCmd{\cmdSend{\varX}{\varY}}{\bcSti, \setenum{\bind{\varCaller}{\pmvA},\bind{\varValue}{0},\bind{\varX}{2},\bind{\varY}{\pmvB}}}{\cmvA}
    \\
    & = \bcSti - \cmvA:2 + \pmvB:2
  \end{align*}
  Let $\bcStii = \bcSti - \cmvA:2 + \pmvB:2$.
  By rule \nrule{[Tx2]}, we obtain 
  $\semBc{\bcB}{\bcSt} = \semTx{\txT[0]}{\bcStii} = \bcStii$.
\end{example}

\section{Additional examples}
\label{sec:tins:examples}

In this~\namecref{sec:tins:examples} we illustrate the expressiveness
of \tins through a series of examples.

\subsection{An extended wallet}
\label{ex:tins:wallet2}

In~\Cref{fig:tins:wallet} we refine the wallet contract in~\Cref{ex:tins:wallet},
by keeping track in the store of the amount of money transferred to each user.

The contract \code{TinyWallet} has two procedures:
\codeFun{init}, which initializes the contract \code{owner}, and
\codeFun{pay}, which transfers \codeVal{amount} units of currency from the
contract to the account \code{dst}.

The procedure \codeFun{init} checks at line~\lineno{4}
if the key \code{owner} is defined;
if not, it means that the contract is still in the initial state
where all keys (except \varBalance) are undefined, and in this case
it binds the key \code{owner} to the \codeVal{sender} of the 
transaction.

The procedure \codeFun{pay} requires at line~\lineno{8} that 
\begin{inlinelist}
\item the caller is the contract owner, 
\item the caller does not transfer any currency along with the call, and
\item the contract balance is enough.
\end{inlinelist}
If any of these conditions does not hold, the procedure throws an exception.
At line~\lineno{10}, if \code{dst} is not bound yet in the store,
then it is set to \codeVal{amount}.
Otherwise, at line~\lineno{11} the old value is incremented 
by \codeVal{amount}.
Finally, line~\lineno{12} transfers \codeVal{amount} units of currency to the recipient.

\begin{figure}[h]
  \begin{center}
    \scalebox{1}{%
\begin{lstlisting}[language=tins,numbers=left,numbersep=5pt,xleftmargin=10pt,classoffset=1,morekeywords={amount,dst},classoffset=2,morekeywords={init,pay}]
contract TinyWallet {

  init() {
    if !owner then throw else owner := sender
  }
    
  pay(amount,dst) {
    if (sender /= ?owner || value /=0 || amount > ?balance) then throw
    else {
      if not !dst then dst := amount
      else dst := ?dst + amount;
      dst $ amount
    }
  }
}
\end{lstlisting}
    }
  \end{center}
  \vspace{-15pt}
  \caption{An extended wallet contract.}
  \label{fig:tins:wallet}
\end{figure}

\begin{figure}[t]
  \centering
  \begin{minipage}[t]{\columnwidth}
\begin{lstlisting}[language=tins,numbers=left,numbersep=5pt,xleftmargin=10pt,classoffset=1,morekeywords={},classoffset=2,morekeywords={init,pay,refund,dispute,resolve,openDispute,closeDispute,fskip}]
contract TinyEscrow
{
  init(x,y) {
    if !buyer then throw
    else { buyer := sender; seller := x; oracle := y }
  }   
  pay() {
    if sender /= ?buyer then throw else ?seller $ ?balance
  }
  refund() {
    if sender /= ?seller then throw else ?buyer $ ?balance
  }
  dispute() {
    if (sender /= ?buyer && sender /= ?seller) then throw
    else ?oracle.openDispute()
  }
}

contract Oracle
{
  init() { isOpen := false }
  
  openDispute() {
    if not ?isOpen then {
      isOpen := true;
      escrow := sender
    }
  }
  closeDispute(z) {
    if sender /= AOracle then throw
    else if ?isOpen {
      fee := ?escrow:?balance * 0.01;
      ?escrow:?buyer $ (?escrow:?balance - ?fee) * z;
      ?escrow:?seller $ ?escrow:?balance;
      ?fee $ AOracle;
      isOpen := false
    }
  }
}
\end{lstlisting}
  \end{minipage}
  \vspace{-15pt}
  \caption{An escrow contract using an oracle.}
  \label{fig:tins:escrow}
\end{figure}

\subsection{An escrow contract}
\label{ex:tins:escrow}

In~\Cref{fig:tins:escrow} we specify in \tins a simple escrow contract,
which allows a \code{buyer} to deposit some funds to the contract
and later authorize their transfer to a \code{seller}.
Further, the \code{seller} can authorize a full refund to the \code{buyer},
in case there is some problem with the purchase.
If \code{buyer} and \code{seller} do not find an agreement,
they can resort to an external authority, which decides how 
the initial deposit is split among them (retaining a fee).

The procedure \codeFun{init} initializes three keys:
\code{buyer} (the sender of the transaction),
\code{seller} and \code{oracle} (passed as parameters).
The guard \code{!buyer} ensures that 
\codeFun{init} can be called at most once.
The procedures \codeFun{pay} and \codeFun{refund} authorize, respectively,
the fund transfer to the \code{seller} or to the \code{buyer};
their guards ensure that a participant cannot authorize a transfer to herself.
Either \code{buyer} and \code{seller} can call \codeFun{dispute},
which in turns calls the procedure \codeFun{openDispute} of the contract at
address \code{oracle}.

A possible contract with this procedure is \code{Oracle} in~\Cref{fig:tins:escrow}:
there, the procedure \codeFun{openDispute} 
just binds the key \code{escrow} to the
address of the contract caller (\code{TinyEscrow}).
The oracle resolves the dispute by calling the procedure
\codeFun{closeDispute}: its parameter \code{z} is the fraction of 
the deposit which goes to the \code{buyer};
1\% of the deposit goes to the oracle as \code{fee}. 
Note that, if \code{buyer} or \code{seller} call
\codeFun{pay} or \codeFun{refund} before the oracle calls
\codeFun{closeDispute}, then the effect of the first four 
instructions within the \code{else} branch of \codeFun{closeDispute} 
is null (since \code{balance} is zero), 
and the invocation just results in the closure of the dispute.

\begin{figure}[t!]
  \begin{center}
    \scalebox{1}{%
\begin{lstlisting}[language=tins,numbers=left,numbersep=5pt,xleftmargin=10pt,classoffset=1,morekeywords={},classoffset=2,morekeywords={init,join,reveal,win,leave}]
contract TinyLottery
{
  init() { nPlayers := 0 }
  
  join(h) {
    if (?nPlayers = 2 || value /= 3) then throw
    else if ?nPlayers = 0
    then { p1 := sender; h1 := h, nPlayers := 1; t0 := Clock:time+1000 }
    else if (h = ?h1) then throw
    else { p2 := sender; h2 := h, nPlayers := 2 }
  }
  leave() {
    if (sender = ?p1 && ?nPlayers = 1 && Clock:time > t0)
    then { ?p1 $ ?balance; nPlayers := 0; }
    else throw
  }
  reveal(s) {
    if (?nPlayers /= 2) then throw
    else if (sender = ?p1 && hash(s) = ?h1 && not !s1) then {s1 := s; ?p1 $ 2}
    else if (sender = ?p2 && hash(s) = ?h2 && not !s2) then {s2 := s; ?p2 $ 2}
    else throw
  }
  win() {
    if (!s1 && !s2)
    then if ((?s1 + ?s2) %2 = 0) then ?p1 $ 2 else ?p2 $ 2
    else if (!s1 && Clock:time > t0) then ?p1 $ 2
    else if (!s2 && Clock:time > t0) then ?p2 $ 2
    else throw
  }
}
\end{lstlisting}
    }
  \end{center}
  \vspace{-15pt}
  \caption{A two-players lottery.}
  \label{fig:tins:lottery}
\end{figure}

\subsection{A two-players lottery}
\label{ex:tins:lottery}

In~\Cref{fig:tins:lottery} we code in \tins a two-players lottery,
inspired by the one in~\cite{Andrychowicz16cacm}.
The players \code{p1} and \code{p2} bet 1 unit of currency each;
additionally, they deposit 2 units of currency as collateral, 
which are used as compensation in case of dishonest behaviour.
The procedure \codeFun{join} allows the players to join the lottery;
the parameter \code{h} is the hash of a secret,
used to implement a timed commitment protocol,
similarly to~\cite{Andrychowicz16cacm}.
The check \code{h = ?h1} at line~\lineno{7} serves to avoid an attack
where the second player replays the same hash of the first one.
The procedure \codeFun{leave} allows the first player to leave the lottery,
if no other player joins before time \code{t0}.
Note that time is provided by an oracle, modelled by the contract \code{Clock}
(not displayed in the figure).
The procedure \codeFun{reveal} allows both players to reveal their secrets:
when this happens, the player redeems her collateral.
Finally, the procedure \codeFun{win} determines the winner of the lottery,
who will collect the bets.
If both players have revealed their secrets, then 
the winner is \code{p1} or \code{p2}, depending on the parity 
of the sum of the secrets.
Otherwise, one player can redeem the bets if
she has revealed her secret and the deadline \code{t0} has passed.

\begin{figure}[t]
  \begin{center}
    \scalebox{1}{%
\begin{lstlisting}[language=tins,numbers=left,numbersep=5pt,xleftmargin=10pt,classoffset=1,morekeywords={},classoffset=2,morekeywords={init,join}]
contract TinyPonzi
{
  init() {
    if !owner then throw
    else { owner := sender;
    	   n := 0;  // total number of investors
	   p := 0   // number of paid investors
    }  
  }
  join() {
    if (value < 1 || not !owner) then throw
    else {
      ?n := (sender,value);
      n := ?n + 1;
      value/10 $ ?owner;
      while (?p < ?n && ?balance >= 2*snd(??p)) do {
        2*snd(??p) $ fst(??p);
        p := ?p + 1
      }
    }
  }
}
\end{lstlisting}
    }
  \end{center}
  \vspace{-15pt}
  \caption{A Ponzi scheme.}
  \label{fig:tins:ponzi}
\end{figure}

\subsection{A Ponzi scheme}
\label{ex:tins:ponzi}
In~\Cref{fig:tins:ponzi} we implement a \emph{Ponzi scheme},
\ie a contract where users invest money, and can redeem their investment (plus interests) 
if enough users invest enough money in the contract afterwards.
In particular, we consider a scheme which pays back users in order of arrival;
this kind of Ponzi schemes gained some popularity in the early stage of Ethereum,
with dozens of different instances.
% ~\cite{Bartoletti17dissecting}.

The procedure \codeFun{init} sets the contract \code{owner}, 
and initializes to $0$ the key \code{n}, which counts the total number of investors,
and \code{p}, which counts the number of investors who have been paid.
The procedure \codeFun{join} allows users to invest money,
and distributes the new investment among all the other users
who have not been paid so far.
The procedure exploits the key-value store to maintain an array of investors.
At line~\lineno{14}, the key \code{?n}
(\ie, the current value bound to \code{n}) 
is bound to a pair, which contains the address of the new investor, and the invested amount.
We use \code{fst} and \code{snd} to access the first and second element of a pair,
respectively.
When a new user joins the scheme, the \code{owner} receives 1/10 of the \code{value} 
transferred along with the call (line~\lineno{16}).
At lines~\lineno{17-19}, the procedure scans the array of unpaid users,
starting from the oldest entry.
As long as the balance is enough, each user receives twice the amount she invested.
Note that \code{?p} denotes the value bound to \code{p}
(\ie, the index of the first unpaid user), 
while \code{??p} denotes the pair $(\varSender,\varValue)$ 
associated to that user.

\section{Conclusions}
\label{sec:tins:comparison}
\label{sec:conclusions}

We have introduced \tins, a minimal core contract calculus inspired by Solidity.
While our calculus is focussed on a single new construct
to call contracts and transfer currency,
other languages have been proposed to capture other peculiar aspects of smart contracts.
Some of these language are domain-specific,
\eg for financial contracts~\cite{Biryukov17wtsc,EgelundMuller17bise}
and for business processes~\cite{LopezPintado19Caterpillar,Tran18bpm},
while some others are more abstract, 
modelling contracts as automata with guarded transitions~\cite{Mavridou19fc,Sergey18scilla}.
Establishing the correctness of the compilation 
from these languages to Solidity would be one of the
possible applications of a bare bone formal model, like our \tins.
Another possible application of a minimal calculus
is the investigation of different styles of semantics,
like \eg denotational and axiomatic semantics.
Further, the study of analysis and optimization techniques for smart contracts
may take advantage of a succinct formalization like ours.

\paragraph{Differences between \tins and Solidity}

Aiming at minimality, \tins simplifies or neglects several features of Solidity.
A first difference is that we do not model a \emph{gas mechanism}.
% used to ensure that all procedure calls terminate.
In Ethereum, when sending a transaction, users deposit into it some crypto-currency,
to be paid to the miner which appends the transaction to the blockchain.
Each computation step performed by the miner consumes part of this deposit;
when the deposit reaches zero, the miner stops executing the transaction.
At this point, all the effects of the transaction
(except the payment to the miner) are rolled back.
Although in \tins we do not model the gas mechanism,
we still ensure that non-terminating calls have an undefined semantics
(see \eg~\Cref{ex:tins:bot}), so that they are rolled back by rule~\nrule{[Tx2]}.
The semantics of \tins could be easily extended with an ``abstract'' gas model,
by associating a cost to instructions and recording the gas consumption in the environment.
However, note that any gas mechanism formalized at the level of abstraction of Solidity
would not faithfully reflect the actual Ethereum gas mechanism,
where the cost of instructions are defined at the EVM bytecode level.
Indeed, compiler optimizations would make it hard to establish a correspondence
between the cost of a piece of Solidity code and the cost of its compiled bytecode.
Still, an abstract gas model could be useful in practice,
\eg to establish upper bounds to the gas consumption of a piece of Solidity code.

A second difference is that our model assumes the set of contracts to be fixed,
while in Ethereum new contracts can be created at run-time.
As a consequence, \tins does not feature constructors that are called
when the contract is created.
Dynamic contract creation could be formalized by extending our model 
with special transactions which extends the mapping $\addrToContr{}$
with the contracts generated at run-time.
Once this is done, adding constructors is standard.

In Ethereum, contracts can implement time constraints by
using the block publication time,
accessible via the variable \code{block.timestamp}.
In \tins we do not record timestamps in the blockchain.
Still, time constraints can be implemented by using oracles,
\ie contracts which allow certain trusted parties to set their keys
(\eg, timestamps), and make them accessible to other contracts
(see \eg the lottery contract in~\Cref{sec:tins:examples}).

In Ethereum, when the procedure name specified in the transaction 
does not match any of the procedures in the contract,
a special unnamed ``fallback’’ procedure (with no arguments) is implicitly invoked.
Extending \tins with this mechanism would be straightforward.
Delegate and internal calls, which we have omitted in \tins,
would be simple to model as well. 

\paragraph{Acknowledgements} 
Massimo Bartoletti and Maurizio Murgia are partially supported 
by Autonomous Region of Sardinia projects \textit{Sardcoin} and \textit{Smart collaborative engineering}.
Letterio Galletta is partially supported by IMT School for Advanced Studies Lucca project \emph{PAI VeriOSS}.
Maurizio Murgia is partially supported by MIUR PON \textit{Distributed Ledgers for Secure Open Communities}.

\bibliographystyle{splncs03}
\bibliography{main}

\begin{thebibliography}{10}
\providecommand{\url}[1]{\texttt{#1}}
\providecommand{\urlprefix}{URL }

\bibitem{solidity}
Solidity documentation. \url{https://solidity.readthedocs.io/en/v0.5.4/} (2019)

\bibitem{Andrychowicz16cacm}
Andrychowicz, M., Dziembowski, S., Malinowski, D., Mazurek, L.: Secure
  multiparty computations on {Bitcoin}. Commun. {ACM}  59(4),  76--84 (2016)

\bibitem{ABC17post}
Atzei, N., Bartoletti, M., Cimoli, T.: A survey of attacks on {Ethereum} smart
  contracts {(SoK)}. In: {POST}. LNCS, vol. 10204, pp. 164--186. Springer
  (2017)

\bibitem{Biryukov17wtsc}
Biryukov, A., Khovratovich, D., Tikhomirov, S.: Findel: Secure derivative
  contracts for {Ethereum}. In: Financial Cryptography Workshops. LNCS, vol.
  10323, pp. 453--467. Springer (2017)

\bibitem{ethereum}
Buterin, V.: {Ethereum}: a next generation smart contract and decentralized
  application platform. \url{https://github.com/ethereum/wiki/wiki/White-Paper}
  (2013)

\bibitem{Crafa19wtsc}
Crafa, S., Pirro, M.D., Zucca, E.: Is {Solidity} solid enough? In: Financial
  Cryptography Workshops (2019)

\bibitem{EgelundMuller17bise}
Egelund{-}M{\"{u}}ller, B., Elsman, M., Henglein, F., Ross, O.: Automated
  execution of financial contracts on blockchains. Business {\&} Information
  Systems Engineering  59(6),  457--467 (2017)

\bibitem{Hildenbrandt18csf}
Hildenbrandt, E., Saxena, M., Rodrigues, N., Zhu, X., Daian, P., Guth, D.,
  Moore, B.M., Park, D., Zhang, Y., Stefanescu, A., Rosu, G.: {KEVM:} {A}
  complete formal semantics of the {Ethereum} {Virtual} {Machine}. In: {IEEE}
  Computer Security Foundations Symposium ({CSF}). pp. 204--217. {IEEE}
  Computer Society (2018)

\bibitem{Igarashi01toplas}
Igarashi, A., Pierce, B.C., Wadler, P.: Featherweight {Java}: a minimal core
  calculus for {Java} and {GJ}. {ACM} Trans. Program. Lang. Syst.  23(3),
  396--450 (2001)

\bibitem{Jiao18arxiv}
Jiao, J., Kan, S., Lin, S., San{\'{a}}n, D., Liu, Y., Sun, J.: Executable
  operational semantics of {Solidity}. CoRR  abs/1804.01295 (2018)

\bibitem{LopezPintado19Caterpillar}
L\'opez-Pintado, O., Garc\'ia-Ba\~nuelos, L., Dumas, M., Weber, I., Ponomarev,
  A.: Caterpillar: A business process execution engine on the {Ethereum}
  blockchain. Software: {Practice} and {Experience}  (2019)

\bibitem{Luu16ccs}
Luu, L., Chu, D.H., Olickel, H., Saxena, P., Hobor, A.: Making smart contracts
  smarter. In: {ACM} {CCS}. pp. 254--269 (2016)

\bibitem{Mavridou19fc}
Mavridou, A., Laszka, A., Stachtiari, E., Dubey, A.: {VeriSolid}:
  Correct-by-design smart contracts for {Ethereum}. In: Financial Cryptography
  and Data Security (2019)

\bibitem{rosu-serbanuta-2010-jlap}
Ro{\c s}u, G., {\c S}erb{\u a}nu{\c t}{\u a}, T.F.: An overview of the {K}
  semantic framework. Journal of Logic and Algebraic Programming  79(6),
  397--434 (2010)

\bibitem{Sergey17wtsc}
Sergey, I., Hobor, A.: A concurrent perspective on smart contracts. In:
  Financial Cryptography Workshops. pp. 478--493 (2017)

\bibitem{Sergey18scilla}
Sergey, I., Kumar, A., Hobor, A.: Scilla: a smart contract intermediate-level
  language. CoRR  abs/1801.00687 (2018)

\bibitem{Tran18bpm}
Tran, A.B., Lu, Q., Weber, I.: Lorikeet: {A} model-driven engineering tool for
  blockchain-based business process execution and asset management. In: {BPM}.
  pp. 56--60 (2018)

\bibitem{Zheng18lolisa}
Yang, Z., Lei, H.: Lolisa: Formal syntax and semantics for a subset of the
  {Solidity} programming language. CoRR  abs/1803.09885 (2018)

\bibitem{Zakrzewski18vstte}
Zakrzewski, J.: Towards verification of {Ethereum} smart contracts: {A}
  formalization of core of {Solidity}. In: {VSTTE}. LNCS, vol. 11294, pp.
  229--247. Springer (2018)

\end{thebibliography}

\end{document}